\documentclass[preprint,12pt]{aastex}
\usepackage{psfig}
\newcommand{\beq}{\begin{equation}}
\newcommand{\eeq}{\end{equation}}
\newcommand{\bea}{\begin{eqnarray}}
\newcommand{\eea}{\end{eqnarray}}
\newcommand{\bean}{\begin{eqnarray*}}
\newcommand{\eean}{\end{eqnarray*}}
\newcommand{\ba}{\begin{array}}
\newcommand{\ea}{\end{array}}
\newcommand{\bml}{\begin{mathletters}}
\newcommand{\eml}{\end{mathletters}}
\newcommand{\rem}[1]{{ }}
\newcommand{\dd}[2]{\frac{{\rm d} #1}{{\rm d} #2}}

\newcommand{\rmmat}[1]{{\hbox{\rm{#1}}}}
\newcommand{\rmscr}[1]{{\hbox{\rm{\scriptsize #1}}}}
\bibpunct[,]{(}{)}{;}{a}{}{,}
\begin{document}
\title{The Synoptic Swift Synergy -- Catching Gamma-Ray Bursts Before They Fly}

\author{Jeremy S. Heyl\altaffilmark{1,2}}
\altaffiltext{1}{Harvard-Smithsonian Center for Astrophysics, Cambridge, 
MA 02138; jheyl@cfa.harvard.edu}
\altaffiltext{2}{Chandra Fellow} 

\begin{abstract}
The advent of large panoramic photometric surveys of the sky offers
the possibly of exploring the association of gamma-ray bursts (GRBs)
with supernovae.  To date, a few gamma-ray bursts have been
connected possibly with supernovae: GRB~980425 -- SN~1998bw,
GRB~011121 -- SN~2001ke, GRB~970228 and GRB~980326.  A combination of
a large detection rate of GRBs and rapid coverage of a large portion
of the sky to faint magnitude limits offers the possibility of
detecting a supernova preceding an associated GRB or at least placing
limits on the rate of association between these two phenomena and the
time delay between them.  This would provide important constraints on
theoretical models for gamma-ray bursts.
\end{abstract}


\section{Introduction}

The gamma-ray burst GRB 021004 was detected by HETE II at 12:06 UT on
the 4th of October 2002 \citep{GCN1565}. Observations after about 9
minutes from the trigger revealed a fading optical transient
\citep{GCN1564}, which was densely sampled in several bands,
especially at early times.  The afterglow of GRB~021004 has shown
several unusual features \citep{GCN1602,GCN1611,Moll02,Bers02b}.
Perhaps its most unique feature was that the field had been
observed shortly before the gamma-ray burst itself was detected.
\citet{GCN1572} give a limiting unfiltered magnitude of 21.4 for
observations on the day before the burst and 22.3 integrated over the
year before the burst.

Astronomy is on the threshold of a new era where large portions of the
sky are surveyed deeply and regularly.  The question arises what is
likelihood of getting photometry of a gamma-ray burst precursor,
specifically if supernovae precede gamma-ray bursts as in the
supranova model \citep{1998ApJ...507L..45V}?  Although the flux upper
limits for GRB~021004 are not strigent enough to constain theoretical
models of gamma-ray bursts, the high-burst localization rate of Swift
combined with the fast sky coverage of the Sloan Digital Sky Survey
(SDSS) and later Pan-STARRS, LSST and the Supernova-Acceleration Probe (SNAP)
could provide important constraints on gamma-ray burst precursors.

During the first and second years of operation of the Swift mission,
SDSS will scan approximately 3000 square degrees (or 7\% of the sky)
each year \citep{SDSS5}.  Over this area it will detect point sources
down to $R\approx 23.2$.  If Swift or subsequent missions are
operational in 2006, the Pan-STARRS program will observe 20,000 square
degrees every four days (or 50\% of the sky) to a limiting magnitude
of $R\approx 24.2$ \citep{PanStarrs}.  Finally, potentially beginning
in 2010, SNAP will cover 15 square degrees every four days with each
observation reaching a limiting magnitude of $R\approx 28$
\citep{SNAP}, coadding observations over a month would go one
magnitude deeper.  The SNAP lensing survey will cover 300 square
degrees over five months to a similar limiting magnitude.

Long gamma-ray bursts are thought to be associated with the collapse
of a massive star, a supernova.  Specifically, in the collapsar model,
the formation of a black hole in the center of the star results in
relativistic jets which pierce the envelope of the star
\citep{1999ApJ...524..262M}.  Along the axis of the jets, the
collapsing star appears as a gamma-ray burst, and the supernova
reaches its peak a few weeks after the GRB.
\citet{1998ApJ...507L..45V} proposed an alternative model in which the
gamma-ray burst accompanies the delayed collapse of a quickly spinning
neutron star which is more massive than the maximum mass of a
non-rotating neutron star.  The neutron star may take several months
or years after the supernova to spin down to the critical frequency
and collapse.

In this {\em Letter}, I will estimate the number of gamma-ray burst
events with photometry which overlaps on the sky but shortly precedes
in epoch from SDSS and other surveys and compare the flux limits with
the expected flux from a supernova which may precede the gamma-ray
burst.

\section{Gamma-Ray Burst Overlap with Future Surveys}

To calculate how often sufficiently deep photometry will precede the
observation of a gamma-ray burst on the sky, several ingredients are
required: a model for the spectral-energy distribution as a function
of time of a supernova associated with a GRB, an estimate of the
luminosity-rate function of GRBs as a function of redshift (${\dot
\phi}(z,L)$), a model for the field of view of the gamma-ray burst
detector ($\Omega_\rmscr{GRB}=2$ for Swift) and its detection
threshold ($P_1$) and the rate of sky coverage of the photometric
program (${\cal R}_\rmscr{photo}$) and its detection threshold
($R_\rmscr{lim}$).  \citet{2001ApJ...548..522P} provide models for
${\dot \phi}(z,L) \equiv R_\rmscr{GRB}(z)\psi(L)$.  The rate of
GRBs, $R_\rmscr{GRB}(z)$, is taken to be proportional to the
star-formation rate, and the luminosity function of GRBs, $\psi(L)$,
is constrained by the BATSE GRB number counts.  The rate of
overlapping photometry is given by the product of the rate of sky
coverage with a integral over the assumed cosmological distribution of
GRBs,
\begin{eqnarray}
\dd{N_\rmscr{overlap}}{t} ( P > P_1, R < R_\rmscr{lim} ) &=& 
\frac{{\cal R}_\rmscr{photo} \Delta t}{4\pi}
 \dd{N_\rmscr{total}}{t} (P>P_1,R<R_\rmscr{lim}) \\
\dd{N_\rmscr{total}}{t} (P>P_1,R<R_\rmscr{lim}) &=&
\frac{\Omega_{\rmscr{GRB}}}{4\pi}
\int_0^{z : R(z) = R_\rmscr{lim}} {\rm d} z
\int_{L(P_1,z)}^\infty {\rm d} L \dd{V(z)}{z} \frac{{\dot
\phi}(z,L)}{1+z}.
\end{eqnarray}
For lack of a better model for the evolution of a supernova associated
with a GRB, I will assume that SN2001ke \citep{2002astro.ph..4234G} is
a prototype for this class, and furthermore that a supernovae
associated with a GRB maintains its peak brightness for a period
$\Delta t = 14 (1+z)$~days in the observer's frame and otherwise it is
undetectable
\citep[see][ for other GRB-associated supernovae]{1999ApJ...521L.111R,1999Natur.401..453B}.
It is reasonable to use the median value of $z$ for
GRBs whose associated supernova are brighter than the magnitude limit
of the particular photometric survey.  However, to be highly conservative, 
I will take $\Delta t=14$~days to calculate the rate of overlap. 

If a survey covers the same area of sky more often than once per 
interval $\Delta t$ such as the SNAP supernovae search and Pan-STARRS,
the rate of sky coverage ${\cal R}_\rmscr{photo}$ should only account
for the first visit in each period $\Delta t$; for example ${\cal
R}_\rmscr{photo}$ for Pan-STARRS is $2 \pi$ per fourteen days.   The 
additional visits during each fortnight do not increase ${\cal
R}_\rmscr{photo}$ but they do allow the survey to probe deeper by
coadding the successive images.

\begin{figure*}
\plottwo{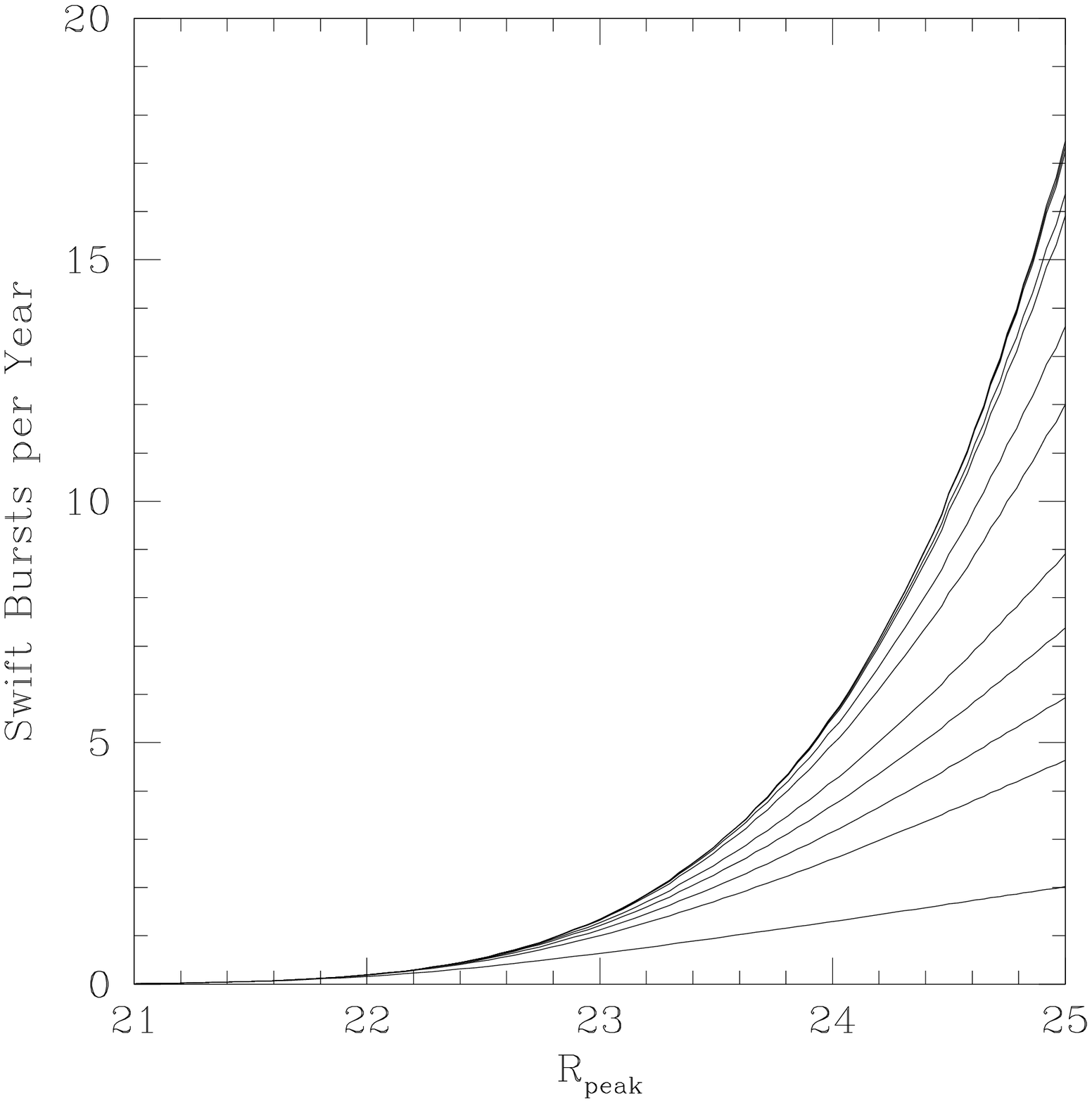}{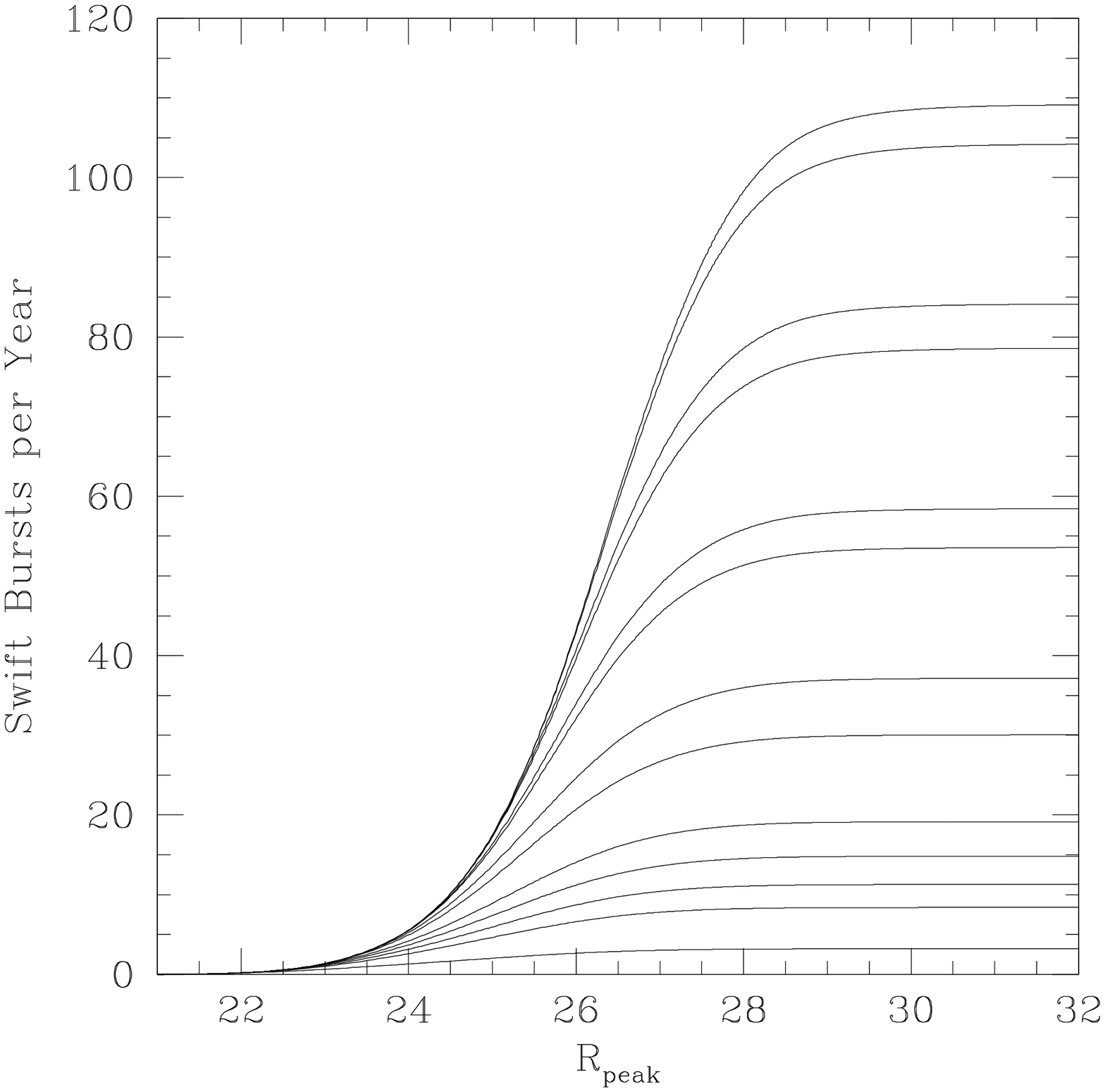}
\caption{The number of GRB-associated supernova brighter than a given
$R$-magnitude.  The lines show the cumulative contribution of GRBs
above given flux limits.  The right panel shows the entire distribution while the 
left panel focusses on the bright end.  From bottom to top, only the supernovae
associated with GRBs whose peak flux is above $10^{0.9}, 10^{0.6},
10^{0.5}, 10^{0.4}, 10^{0.3}, 10^{0.1}, 1,
10^{0.2},10^{-0.25},10^{-0.45},10^{-0.5},10^{-0.7}$ and
$10^{-0.75}$~photons per square centimeter per second.  See
\citet{2001ApJ...548..522P} for further details.}
\label{fig:number}
\end{figure*}

According to the original supranova model
\citep{1998ApJ...507L..45V}, the supernova may reach its peak at any
time up to several years before the GRB, so this calculation
implicitly assumes that both the GRB survey and the photometric survey
will be operating at the appropriate times.  $L(P_1,z)$ is the
luminosity of a GRB at a redshift $z$ which is detected at a
count-rate of $P_1$, and $R(z)$ is the $R$-band apparent magnitude of
a GRB-associated supernovae at a redshift $z$.  Both of these
functions include the $k-$correction \citep{Hogg99} and assume the
cosmographic parameters, $\Omega_M=0.3$ and $\Omega_\Lambda=0.7$.
Fig.~\ref{fig:number} plots the number of GRBs detected by Swift per
year whose associated supernova would be brighter at its peak than a
particular $R$-band magnitude.

The results shown in Fig.~\ref{fig:number} assume the SF1 model of 
\citet{2001ApJ...548..522P}.  This model provides a conservative
lower limit for the overlap.  It predicts that Swift will localize
about 110 bursts per year --- the more generous estimates range up to
300 bursts per year \citep{Swift}.  Furthermore, this model predicts
that the bursts detected will be a higher redshifts than other models,
so the accompanying supernovae will be fainter and more difficult to
detect. 

Table~\ref{tab:survey} gives the overlap rate between various
photometric surveys and the Swift GRB localization mission.  What is
striking is that the shallow but wide Pan-STARRS and LSST surveys will
perform much better than any of the other surveys.  Furthermore, if
supernovae precede GRBs, Pan-STARRS and LSST each will detect nearly ten
GRB-associated supernovae per year.  If it finds none, it would place
severe constraints on the supranova model for GRBs.  It must be
emphasized that this rate of overlap is extremely conservative.  It
assumes a low Swift burst localization rate and a distribution of GRBs
skewed to high redshift (therefore, faint assocated supernovae).  The
actual rate of overlap will probably be higher if both programs
operate simultaneously.  Furthermore, Swift will generate a catalog of
burst positions and redshifts.  One will be able to cross-correlate
{\em a posteriori} this catalog with earlier Pan-STARRS or LSST observations
and exclude the appearance of transients to $R \approx 25$ over a wide
range of epochs preceding the burst yielding definitive constraints on
GRB-progenitors independent of assumptions about the GRB luminosity
function and its evolution.

This calculation of the overlap rate assumes that either the GRB
localization program or the photometric survey studies random portions
of the sky.  In fact both the Swift mission and all of the photometric
surveys avoid studying the region of the sky near the sun.  Although
the average rate of overlap over a year in given by the formulae above
and the values in the tables, the chance of detecting the supernova
associated with GRB is somewhat higher than average if the supernova
precedes the GRB by less than three months or between nine and fifteen
months.  If the supernova precedes the GRB by six to nine months, the
chance of detecting it is somewhat lower than average.  However, this
seasonal variation is smaller than the uncertainities in the GRB luminosity
function.
\begin{deluxetable}{l|cccccc}
\tablecaption{Present and Future Large-Scale Photometric Surveys
\label{tab:survey}
}
\tablehead{\colhead{Survey} & \colhead{$R_\rmscr{lim}$} & 
\colhead{$z_\rmscr{max}$} &
\colhead{$z_\rmscr{med}$} &
\colhead{${\cal R}_\rmscr{photo}\Delta t$} & 
\colhead{$d N_\rmscr{total}/dt [\rmmat{y}^{-1}]$} & 
\colhead{$d N_\rmscr{overlap}/dt [\rmmat{y}^{-1}]$ } }
\startdata
SDSS                & 23.2 & 0.56 & 0.48 &  0.035  & 1.9  & 0.0052 \\
Pan-STARSS (single) & 24.2 & 0.78 & 0.66 &  6.3    & 7.1  & 3.6 \\
Pan-STARSS (coadded)& 25.0 & 1.00 & 0.83 &  6.3    & 17.  & 8.8  \\
LSST (single)       & 24.5 & 0.86 & 0.72 &  6.3    & 10.  & 5.1  \\
LSST (coadded)      & 25.1 & 1.04 & 0.85 &  6.3    & 19.  & 9.8  \\
SNAP SN (single)    & 28.0 & 2.59 & 1.45 &  0.0046 & 98. & 0.036  \\
SNAP SN (coadded)   & 28.8 & 3.35 & 1.50 &  0.0046 & 110. & 0.039 \\
SNAP lensing        & 28.0 & 2.59 & 1.45 &  0.0091 & 98. & 0.071  \\
\enddata
\end{deluxetable}

\section{Discussion}

The philosophy employed for finding gamma-ray burst precursors is
somewhat different that what is necessary for finding supernovae or
microlensing events.  Because the precursors will be sought after the
gamma-ray burst is detected and localized, it is not necessary to have
more than one epoch of data from the particular region of sky before
the burst.  Even a single epoch would yield important constraints.
Furthermore, unless the cadence of the observations is sufficiently
low (no more than biweekly), the repeated observations of the same
patch of sky do not improve the chances of catching a precursor
(unless one coadds the data to probe deeper), because supernova
typically evolve over the course of weeks.  Consequently, although
supernova and microlensing surveys have a large data rate of high
quality photometry, because of their relative lack of sky coverage and
depth they do not contribute much to the detection rate of precursors.
From another point of view, only a small fraction ($<10^{-4}$) of
supernovae result in GRBs directed toward us, so one would typically
have to find at least $10^4 (4\pi/\Omega_\rmscr{GRB})$ supernovae in a
blind search before finding a single GRB-associated supernova.

The best bets are the large deep wide surveys of the sky. SDSS is the
prototype and Pan-STARRS and LSST should deliver results.  There is a
small possibility that SDSS will catch a supernova before a gamma-ray
burst providing important evidence for the supranova model for
gamma-ray bursts (it may have done so already).  Pan-STARRS or LSST if it
overlaps with a high-locatization-rate GRB mission such as Swift will
be able to provide important constraints on gamma-ray-burst models.
Specifically, it will be able to exclude the possibility that GRBs
follow supernovae within a year.

\acknowledgements 

I would like to acknowledge useful discussions with Kris Stanek,
Robert Lupton and Bob Kirshner and helpful suggestions from the
anonymous referee.  I was supported by the Chandra Postdoctoral
Fellowship Award \# PF0-10015 issued by the Chandra X-ray Observatory
Center, which is operated by the Smithsonian Astrophysical Observatory
for and on behalf of NASA under contract NAS8-39073.

\bibliographystyle{apj}
\bibliography{mine,021004,cosmo,gr}

\end{document}